\documentclass[amsmath,amssymb,preprint]{revtex4-2}

\pdfoutput=1

\usepackage[dvips,final]{graphicx}
\usepackage{amsmath}
\usepackage[dvipsnames,usenames]{color}
\usepackage[english]{babel}
\usepackage{subfigure}  

\begin{document}

\title{Evaluating the ratio of the exclusive vector meson photoproduction to inclusive hadron/jet production cross section in ultraperipheral heavy ion collisions}

\author{Joao Vitor C. Lovato$^{1}$}
\email{joaovitorcl1000@gmail.com}

\author{Edgar Huayra$^{1}$}
\email{yuberth022@gmail.com}


\author{Magno V.T. Machado$^{2}$}
\email{magnus@if.ufrgs.br}

\affiliation{
\\
{$^1$\sl Departamento de F\'isica, CFM, Universidade Federal 
de Santa Catarina, C.P. 5064, CEP 88035-972, Florian\'opolis, 
SC, Brazil \\
$^2$\sl High Energy Physics Phenomenology Group, GFPAE. Institute of Physics, Federal University of Rio Grande do Sul (UFRGS) Caixa Postal 15051, CEP 91501-970, Porto Alegre, RS, Brazil
}
}

\begin{abstract}
\vspace{0.5cm}

Using the QCD color dipole picture to study exclusive vector meson and inclusive jet/open meson photoproduction, we calculate the ratio of elastic meson production to inclusive hadron production cross sections for ultraperipheral heavy ion collisions. Predictions are evaluated for run 4 of the Large Hadron Collider in proton-nucleus ($pA$) and nucleus-nucleus ($AA$) collisions. The dependencies of the ratio on jet/hadron transverse momentum and atomic number are investigated. The double ratio $R_{\mathrm{UPC}}$ for $AA$ over $pA$ collisions is also computed, which has been previously proposed as a new observable probing parton saturation physics.

\end{abstract}
\maketitle

\section{Introduction}
\label{Sect:intro}

High--energy physics is continually advancing as particle accelerators enable us to investigate increasingly extreme kinematic regions. At this regime, many Quantum Chromodynamics (QCD) phenomena appear, such as the consequences of DGLAP~\cite{Gribov:1972ri, Altarelli:1977zs, Dokshitzer:1977sg} and BFKL~\cite{Fadin:1975cb, Lipatov:1976zz, Kuraev:1977fs, Balitsky:1978ic, Fadin:1998py} evolution of partons. Another intriguing phenomenon, associated with small values of Bjorken $x$, emerges. Such a low $x$ region leads to an increase in parton densities within hadron and nuclear wave functions. However, this rise in parton densities cannot continue indefinitely. As the density reaches sufficiently high levels, the probability of interactions between partons increases, resulting in nonlinear effects related to processes such as parton merging or recombination. These interactions moderate the further growth of the parton distributions, effectively slowing down the increase as $x$ decreases. This phenomenon is known as saturation (see reviews~\cite{Mueller:2001fv, Blaizot:2011pa, Jalilian-Marian:2005ccm, Gelis:2010nm, Munier:2009pc, Weigert:2005us, Morreale:2021pnn}). The state where the rate of parton mergers balances the rate of parton splittings is referred to as the parton saturation regime. This concept is important in studying high--energy nuclear physics because it represents a limit beyond which the usual perturbative approaches to QCD may no longer be valid. 

The understanding of saturation behavior remains one of the main challenges in contemporary hadron physics, and a major focus is to unambiguously identify signatures of parton saturation in experimental data~\cite{Morreale:2021pnn}. This task is critical for experiments at current and future facilities, such as the planned Electron--Ion Collider (EIC)~\cite{Accardi:2012qut, AbdulKhalek:2021gbh} and the Large Hadron Electron Collider (LHeC)~\cite{LHeC:2020van, Ahmadova:2025vzd}, which aim to explore the structure of matter at high energies. The EIC and LHeC experimental programs include a significant focus on studying parton saturation and other related phenomena. Additionally, the Large Hadron Collider (LHC) can help identify saturation signatures by using photon–hadron collisions to constrain hadron structure at low–$x$~\cite{Hentschinski:2022xnd}. Moreover, the low–$x$ geometric scaling phenomenon, which has a straightforward interpretation within parton saturation physics, is investigated in soft observables measured at the LHC~\cite{Baldenegro:2022xrj, Baldenegro:2024vgg, Peschanski:2024tlr}. 

One approach to investigating saturation signatures is to relate them to other phenomena that allow us to probe saturation effects, such as the black disk limit. This behavior is also associated with larger parton densities, since, in this regime, the target (such as a proton or nucleus) becomes progressively more ``opaque'' to incoming particles. This implies that an incoming particle is increasingly likely to interact with one of the partons in the target. In the extreme case where the target is completely opaque, it behaves like a black disk---an idealized object in which any particle that hits it will interact and be either absorbed or scattered. However, the interaction probability cannot increase indefinitely due to unitarity, which imposes an upper bound on cross--sections. Specifically, the elastic scattering amplitudes must satisfy the condition that the total probability of all possible outcomes (scattering and non--scattering) sums to 1. In the black disk limit, this condition implies that $50 \%$ of the total cross--section corresponds to elastic scattering (the probability of deflection), while the other $50 \%$ corresponds to inelastic interactions (the probability of absorption or interaction). The black disk limit is closely connected to saturation physics since once the system reaches saturation, it effectively behaves as a black disk, where further increases in parton densities do not significantly affect the interaction probabilities. Then, away from the black disk limit, the elastic--to--total cross--section ratio is small, $\sigma_{\text{el}}/\sigma_{\text{total}} \ll 1$, while saturation is associated with a ratio approaching $\sim \frac{1}{2}$. Based on this idea, one saturation signal was proposed at the EIC~\cite{Accardi:2012qut}. Such an observable is the fraction of diffractive DIS with respect to the total DIS cross section in electron--nucleus $eA$--collisions. 

The Kovchegov--Sun--Tu approach~\cite{Kovchegov:2023bvy} proposes a similar ratio that is also sensitive to saturation effects and can be measured in the ultraperipheral collisions (UPCs) in the LHC and RHIC. The study of UPCs is particularly interesting due to their clean signal for photon--mediated interactions~\cite{Baur:2001jj, Greiner:1992fz, Bertulani:2005ru, Baltz:2007kq, Contreras:2015dqa, Klein:2017nqo, Klein:2020fmr}. In UPCs, the elastic cross section for vector meson production is quadratic in the dipole amplitude $N$, whereas the inclusive particle production cross section is linear in $N$. This difference in sensitivity to the dipole amplitude makes the new ratio $R_{\text{UPC}}$ sensitive to saturation effects and the system's approach to the black disk limit. The new ratio compares processes with different dependencies on the saturation scale, making it a more effective tool for studying the approach to the black disk limit in UPCs. It bypasses the difficulty of measuring diffractive DIS cross sections in electron--ion collisions by focusing on photoproduction in UPCs, which is more accessible in these types of particle reactions. Additionally, the ratio is particularly useful for studying $A$--scaling (dependence on the nucleus size $A$), which could indicate the presence of saturation.

The authors of~\cite{Kovchegov:2023bvy} demonstrate that the $A$--scaling behavior of the double ratio $R_{\rm{UPC}}$ defined as:
\begin{eqnarray}\label{eq:R_UPC}
    R_{\mathrm{UPC}} = \frac{\left[ \sigma^{VM}_{\text{el}}/\left(d\sigma^{\text{jet}}_{\text{inc}}/d^2 p_{\perp} \right) \right]_{AA} }{\left[\sigma^{VM}_{\text{el}}/\left( d \sigma^{\text{jet}}_{\text{inc}}/d^2 p_{\perp} \right) \right]_{pA}},
\end{eqnarray}
differs between the saturated and non--saturated regimes. Specifically, the $A$--dependence of $R_{\rm{UPC}}$ varies according to the size of the produced vector meson---being different for a small meson like the $J/\psi$ and a large meson like the $\rho$. Additionally, it depends on whether the transverse momentum $p_\perp$ of the produced hadron or jet is greater or smaller than the saturation scale $Q_s$. These dependencies reflect how the system transitions into or out of the saturation regime. Approximate analytical predictions were proposed, assuming a fixed center--of--mass (c.o.m) energy of the photon--nucleus/nucleon system. 

The aim of the present work is to implement realistic numerical calculations of the observable $R_{\rm{UPC}}$ for the LHC Run 4~\cite{Dainese:2019rgk} using the state of the art in small--$x$ QCD phenomenology. The main theoretical uncertainties for the double ratio are discussed. Moreover, predictions for the ratio $\sigma^{VM}_{\text{el}}/(d\sigma^{\text{jet}}_{\text{inc}}/d^2 p_{\perp})$ are presented in $pA$ and $AA$ collisions separately.  The hadronization effects in the inclusive production of open mesons in the final state are also investigated. The experimental measurement of this process is now a reality after the CMS Collaboration determination of the inclusive $D$ mesons photoproduction in UPCs \cite{CMS:2024ayq}. The structure of this paper is organized as follows: In Section II, we provide a brief overview of the theoretical framework used for vector meson and single inclusive jet photoproduction, which has been extensively discussed in the literature. In Section III, we present and discuss our results for the predicted saturation ratio under LHC Run 4 kinematics, including individual ratios for proton--nucleus and nucleus--nucleus systems. Additionally, we examine the broader implications of our findings. Finally, in Section IV, we conclude with a summary of the key results and their significance.

\section{Theoretical framework}
\label{Sect:formalism}

The investigation of photon--induced interactions in hadron collisions has garnered substantial interest in recent years~\cite{Klein:2020fmr}, primarily due to the potential to improve our understanding of QCD dynamics at high energies~\cite{Morreale:2021pnn}. In the literature, this process is referred to as ultraperipheral collisions (UPCs) and is characterized by large impact parameters, where the interacting nuclei remain sufficiently distant to avoid direct hadron interactions. In $pA$ or $AA$ systems, UPCs occur by the interaction of a quasi--real photon (photons with very small virtuality, namely $Q^2 \approx 0$), emitted by one nucleus, with the other nucleus or proton as the target~\cite{Baur:2001jj, Greiner:1992fz, Bertulani:2005ru, Baltz:2007kq, Contreras:2015dqa, Klein:2017nqo, Klein:2020fmr}. At high energies (small $x$), the dominant process can be described by the color dipole model, where the photon fluctuates into a $q\bar{q}$ pair, which then interacts with the target nucleus~\cite{Nikolaev:1990ja, Nikolaev:1991et, Mueller:1993rr}. 

Before presenting our results, we first discuss the theoretical framework underlying the Kovchegov--Sun--Tu ratio. In particular, we summarize the main assumptions and approximations that enter the formalism, including the treatment of the photon wave function, the dipole--nucleus scattering amplitude, and the relevant kinematic variables that govern the high--energy behavior of the process.

\subsection{Elastic photoproduction off the target}

One of the most widely studied processes in UPC interactions is the exclusive production of a vector meson, which has been extensively measured in proton--proton ($pp$), proton--lead ($p$Pb), and lead--lead (PbPb) collisions~\cite{Tlusty:2025bsh, Ragoni:2024dlh, CMS:2024krd, McNulty:2024qdr}. Within the dipole framework, the virtual photon splits into a quark--antiquark pair $q\bar{q}$, which then undergoes elastic scattering off the target, probing its internal structure through strong interactions. Notably, both nuclei remain intact in the final state, as no direct hadron interactions take place, and the cross section is directly proportional to the square of the gluon distribution in the collinear formalism~\cite{Guzey:2013qza, Kryshen:2023bxy, Schenke:2024gnj}, making it a valuable probe for gluon saturation effects~\cite{Schenke:2024gnj}.

At leading order (LO) in dipole formalism, the elastic cross section integrated over the impact parameter for the process $\gamma p \rightarrow V p$ to produce a vector meson $V$ can be approximated by taking the average over the square modulus of the interaction amplitude~\cite{Nemchik:1996cw, Hufner:2000jb}:
\begin{eqnarray}
   \sigma_{\text{el}}^{\gamma p \to V p}(x, Q^2 \approx 0) &=& \frac{1}{16 \pi B_V}  \left| \int_0^1 \frac{dz}{z(1-z)} \int  \frac{d^2 r_\perp}{4\pi} \left[\Psi_{V}^* \Psi^{\gamma \to q\bar{q}}\right]_{T}(\mathbf{r}_\perp, z)\, \sigma_{\text{dip}, \, p}(x, \mathbf{r}_\perp) \right|^2. \nonumber \\
   & &
   \label{eq:elastic}
\end{eqnarray}
Here, the $B_V$ is the elastic slope parameter typically fitted to the exclusive quarkonia electroproduction data available from the HERA collider. The $z$ is the longitudinal momentum fraction carried by the quark in the dipole (with the antiquark carrying $1 - z$), and $r_\perp \equiv |\mathbf{r}_\perp|$ denotes the transverse separation of the $q\bar{q}$ pair. The function $\Psi^{\gamma \to q\bar{q}}_{T}$ denotes the probability amplitude for the incoming virtual photon to fluctuate into a quark--antiquark pair with transverse polarization. In turn, $\Psi^V$ represents the probability amplitude for the dipole to hadronize into the final--state vector meson.

In the nuclear case, we restrict our analysis to coherent photoproduction, $\gamma A \to V A$, in which the nucleus remains intact~\cite{Ivanov:2002kc}:
\begin{eqnarray}
   \sigma_{\text{el}}^{\gamma A \to V A}(x, Q^2 \approx 0) &=& \mathcal{K}_V \int d^2 s_\perp \, \left| \int_0^1 \frac{dz}{z(1-z)} \int  \frac{d^2 r_\perp}{4\pi} \left[\Psi_{V}^* \Psi^{\gamma \to q\bar{q}}\right]_{T}(\mathbf{r}_\perp, z)\, N(x, \mathbf{r}_\perp, \mathbf{s}_\perp) \right|^2. \nonumber \\
   & &
   \label{eq:nuclearelastic}
\end{eqnarray}
The $N$ in Eq.~\ref{eq:nuclearelastic} is the imaginary part of the forward scattering amplitude $N$. This quantity, which characterizes the scattering between the dipole and the target $A$, is related to the dipole cross section by $\sigma_{\text{dip},\, A} =  2 \int d^2 s_\perp \, N$. Usually it is assumed to be independent of the $\mathbf{r}_\perp$ direction, $N(x, \mathbf{r}_\perp, \mathbf{s}_\perp) \equiv N(x, r_\perp, s_\perp)$, valid in the quasi--classical approximation for a large nucleus. This quantity is commonly parameterized using theory--based fits~\cite{Motyka:2008jk}. The normalization factor $\mathcal{K}_{V}$ accounts for corrections due to the real part of the amplitude and to skewedness effects~\cite{Bronzan:1974jh, Shuvaev:1999ce, Goncalves:2004bp}, not included here. 

\begin{table}[t]
     \setlength{\tabcolsep}{10pt} 
    \renewcommand{\arraystretch}{1.2} 
    \centering
    \begin{tabular}{|c || c | c | c | c|| c | c|} \hline \hline
        Meson & $Z_f$ & $M_V/$ GeV & $B_V/$ $\text{GeV}^{-2}$ & $\mathcal{K}_{V}$ & $\mathcal{N}_T$ & $R^2/$ $\text{GeV}^{-2}$ \\ \hline
         $J/\psi$  & $2/3$ & $3.097$ & $4.57$ & $1/1.1$ & $0.602$ & $2.3$ \\ \hline
        $\rho$ & $1/\sqrt{2}$ & $0.775$ & $9.00$ & $1/0.7$ &  $0.909$ & $12.9$ \\ \hline
    \end{tabular}
    \caption{Parameters for the Boosted-Gaussian wave functions of $J/\psi$ and $\rho$ vector mesons, calculated following the prescription described in \cite{Kowalski:2006hc}, using the current PDG values for quark and meson masses~\cite{ParticleDataGroup:2022pth}.}
    \label{tab:parametersmeson}
\end{table}

Using the same approach of Kovchegov--Sun--Tu, we work with the Boosted--Gaussian model~\cite{Nemchik:1996cw, Forshaw:2003ki} for the scalar part of the vector meson wave function. Other approaches are possible, like the models based on holography~\cite{Forshaw:2012im, Ahmady:2016ujw} and other Gaussian--like models~\cite{Nemchik:1996cw, Dosch:1996ss}. However, for our purpose, once we treat a ratio, this dependence is weak. Then, by neglecting the longitudinal components, the elastic cross section for a transversely polarized photon in Eq.~\ref{eq:nuclearelastic} becomes~\cite{Kovchegov:2023bvy}:
\begin{eqnarray} \label{eq:elasticfinalcrosssection}
\sigma^{\gamma A \to VA}_{\text{el}}(x, Q^2 \approx 0) &=& \mathcal{K}_{V} \frac{\alpha_{\text{em}} Z_f^2}{\pi} N_c^2 \mathcal{N}_T^2 e^{m_f^2 R^2} \int_0^1 dz \, dz' \int_0^\infty dr_\perp \, dr_\perp' \, r_\perp \, r_\perp' \\
& & \nonumber \\
&\times& \left\{ \frac{4}{R^2} z(1 - z) \left[ z^2 + (1 - z)^2 \right] r_\perp a_f K_1 (r_\perp a_f) + m_f^2 K_0 (r_\perp a_f) \right\} \nonumber \\
&\times& \left\{ \frac{4}{R^2} z'(1 - z') \left[ z'^2 + (1 - z')^2 \right] r_\perp' a_f' K_1 (r_\perp' a_f') + m_f^2 K_0 (r_\perp' a_f') \right\} \nonumber \\
 & & \nonumber \\
 &\times& \exp \left[-\frac{2z (1 - z) r_\perp^2}{R^2} - \frac{m_f^2 R^2}{8z(1 - z)} \right] \exp \left[-\frac{2z' (1 - z') r_\perp'^2}{R^2} - \frac{m_f^2 R^2}{8z'(1 - z')} \right] \nonumber \\
 & & \nonumber \\
&\times& \int d^2 s_\perp\, N(x, r_\perp, s_\perp) N(x, r_\perp', s_\perp), \nonumber
\end{eqnarray}
with a similar expression for $\sigma^{\gamma p \to Vp}_{\text{el}}$. 

In this equation, the $K_0$ and $K_1$ denote the modified Bessel functions of the second kind, while $m_f$ and $Z_f$ represent the quark mass and charge fraction for flavor $f$ ($= u, d, s$ and $c$), respectively. The quantity $a_f^2 = z(1 - z)Q^2 + m_f^2$ combines the photon virtuality and the quark mass. Here, $N_c = 3$ is the number of QCD colors, and $\alpha_{\text{em}} = e^2/(4\pi)$ denotes the electromagnetic coupling constant. Both the normalization constant $\mathcal{N}_T$ and the parameter $R$, which controls the width of the wave function, originate from the Boosted--Gaussian ansatz used to model the scalar part of the vector meson wave function. The values of all relevant parameters are summarized in Tab.~\ref{tab:parametersmeson}. In the c.o.m frame of the colliding particles, the produced vector meson state $V$ is characterized by its mass $M_V$ and $\gamma N$--system energy $W_{\gamma N}$. Then, the Bjorken variable $x$ for quasi--real photons ($Q^2 \approx 0$) is approximately $x \approx M_V^2/W_{\gamma N}^2$. 

\subsection{Single inclusive jet photoproduction}

Another approach to probing QCD dynamics within the dipole formalism is through photon--induced processes, where one of the incident particles (hadron or nucleus) fragments. These are referred to as (semi–)inclusive processes, with notable examples including heavy--quark production and inclusive dijet photoproduction in hadron collisions (see, for instance, Ref.~\cite{Guzey:2018dlm, Klein:2002wm, Goncalves:2004dn}). As in vector meson photoproduction, the inclusive single quark production cross section is computed using $\gamma A$ scattering via dipole formalism~\cite{Mueller:1999wm, Kovchegov:2015zha, Floter:2007xv}. However, a key difference in this scenario is that the interaction is no longer restricted to a color--singlet exchange, implying the target proton or nucleus may fragment. Additionally, the quark--antiquark ($q\bar{q}$) pair generated from the virtual photon splitting does not recombine into a vector meson--instead, the quark and antiquark fragment independently into hadrons or jets. The quark production cross section is specifically calculated for the produced quark, and under the eikonal approximation, it is identical to that of the antiquark.

The cross section for single jet photoproduction can be expressed in terms of the photon wave function $\Psi$. In particular,  the transverse momentum distribution for a $p A$ and $AA$ scattering of a single $f$--jet with transverse momentum $p_\perp$ will be~\cite{Mueller:1999wm, Kovchegov:2015zha, Floter:2007xv}:
\begin{eqnarray}
   \frac{d\sigma^{\gamma A \to f X}}{d^2 p_\perp}(x, p_\perp,Q^2 \approx 0) &=& \frac{4 N_c \alpha_{\text{em}} Z_f^2}{(2\pi)^2} \int_0^1 dz  \left\{m_f^2\left[\frac{\mathcal{I}_1}{p_{\perp}^2 + a_f^2}-\frac{\mathcal{I}_2}{4a_f} \right] \right. \nonumber \\
   &+& \left. \left[z^2 + (1 - z)^2 \right] \left[ \frac{a_f p_\perp \mathcal{I}_3 }{p_{\perp}^2 + a_f^2}  -\frac{\mathcal{I}_1}{2} + \frac{a_f \mathcal{I}_2}{4}\right]  \right\},  
   \label{eq:f-jet}
\end{eqnarray}
where $A = 1$ for proton and $A > 1$ for nucleus. The $a_f^2 = z (1-z)Q^2+m_f^2$, the color number $N_c$, the quark mass $m_f$, the quark charge $Z_f$, and the $\alpha_{\text{em}}$ fine--structure constant were defined previously. The auxiliary functions that involve Bessel functions are
\begin{subequations} \label{integrals}  
    \begin{align}
\mathcal{I}_1& = \int_{0}^{\infty} dr_\perp \,r_\perp\, J_0(p_\perp r_\perp)K_0(a_f r_\perp)\,\sigma_{\text{dip},\, A}(\bar{x},r_\perp), \\
\mathcal{I}_2& = \int_{0}^{\infty} dr_\perp \,r_\perp^2\, J_0(p_\perp r_\perp)K_1(a_f r_\perp)\,\sigma_{\text{dip},\, A}(\bar{x}, r_\perp), \\
\mathcal{I}_3& = \int_{0}^{\infty} dr_\perp \,r_\perp\, J_1(p_\perp r_\perp)K_1(a_f r_\perp)\,\sigma_{\text{dip},\, A}(\bar{x}, r_\perp).
    \end{align}
\end{subequations}
Here, the Bjorken variable is $\bar{x} \approx M_\perp^2/W_{\gamma N}^2$, where $M_\perp := \sqrt{p_\perp^2+ m_f^2}$ is the transverse mass of the jet, $J_0$ and $J_1$ are the modified Bessel functions of the first kind, and the other variables are the same as those defined previously in the vector meson discussion. 

In UPCs, the final state of this semi--inclusive process is characterized by a single rapidity gap, indicative of the photon exchange, and an intact hadron that acts as the photon source. Inclusive and exclusive processes differ in two key aspects: first, the cross section for inclusive processes is proportional to the first power of the gluon distribution in the collinear formalism, whereas for exclusive processes, it scales with the square of the gluon distribution, making cross sections of the inclusive process generally an order of magnitude larger. Second, inclusive processes are more challenging to isolate experimentally due to the significant background noise compared to exclusive processes.

\subsection{Integrated cross sections and photon flux}

As mentioned previously, the impact parameter satisfies the condition $b > R_{A} + R_{p,\, A}$ for UPC collisions, where $R_{p,\, A}$ denotes the radius of a proton or a nucleus. This is a necessary condition to ensure that photon--induced interactions dominate the process. In the equivalent photon approximation (EPA)~\cite{Fermi:1924tc,vonWeizsacker:1934nji, Williams:1934ad}, ultra--relativistic nuclei act as sources of quasi--real photons that are collinear with them. Consequently, the square of the c.o.m energy of the photon--nucleon system is given by $W_{\gamma N}^2 = 4E_{\gamma} E_N = 2 \omega \sqrt{s_{NN}}$, where $\sqrt{s_{NN}} = \sqrt{Z_A/A} \, \sqrt{s_{pp}}$ for $pA$ collisions and $\sqrt{s_{NN}} = (Z_A/A) \, \sqrt{s_{pp}}$ for $AA$ collisions represents the c.o.m energy per nucleon--nucleon pair, $\sqrt{s_{pp}}$ is the c.o.m proton--proton collision, and $Z_A$ and $A$ represent, respectively, the atomic and mass number of the nucleus. In this scenario, the projectile is a nucleon within the incoming nucleus, while the target is a nucleon within the target. Then, the photon energy, denoted by $\omega$, is related to these variables and final--state--particle rapidity $y$ by
\begin{eqnarray} \label{eq:omega_kinematics}
    \omega_{\text{el}} = \frac{W_{\gamma N}^2}{2\sqrt{s_{NN}}} \approx \frac{M_V}{2} e^{y}, \quad \omega_{\text{inc}} =  \frac{W_{\gamma N}^2}{2\sqrt{s_{NN}}} \approx \frac{M_\perp}{2 z} e^{y}.
\end{eqnarray}

For $pA$ collisions, we only consider the case where the nucleus is the source. However, for $AA$ collision, in order to perform numerical calculations of the differential cross sections, it is crucial to account for the fact that the photon can be emitted by either of the nuclei. This is handled by incorporating the substitution $y \to -y$ in rapidity, ensuring symmetry in the process. However, when the colliding nuclei are identical, this effect can be accounted for by simply multiplying the cross section by a factor of 2. The differential cross sections for proton--nucleus and nucleus--nucleus collisions are then computed by approximating the transverse position of the photon to the impact parameter of the collision, $b \approx b_\gamma$, while neglecting the contribution of the proton as a quasi-real photon source~\cite{Baur:2001jj, Greiner:1992fz, Bertulani:2005ru, Baltz:2007kq, Contreras:2015dqa, Klein:2017nqo, Klein:2020fmr}:
\begin{subequations} \label{mesonspectrum}  
    \begin{align}
  \frac{d \sigma^{pA \to pVA}_{\text{el}}}{d \omega}(\omega, Q^2)
    &\approx n_\gamma^{pA}(\omega)\  \sigma^{\gamma A \to VA}(\omega, Q^2), \\
    \frac{d \sigma^{AA \to AVA}_{\text{el}}}{d \omega}(\omega, Q^2)
    &\approx 2\ n_\gamma^{AA}(\omega)\ \sigma^{\gamma A \to VA}(\omega, Q^2),
    \end{align}
\end{subequations}
for elastic cross sections and, for jets $p_\perp$ distributions, we have
\begin{subequations} \label{jetspectrum}  
    \begin{align}
    \frac{d\sigma^{pA \to f p X}}{d\omega d^2 p_\perp}(\omega, p_\perp, Q^2) &\approx n_\gamma^{pA} (\omega) \ \frac{d\sigma^{\gamma p \to f X}}{d^2 p_\perp}(\omega, p_\perp, Q^2), \\
    \frac{d\sigma^{AA \to f A X}}{d\omega d^2 p_\perp}(\omega, p_\perp, Q^2) &\approx 2 \  n_\gamma^{AA} (\omega)\  \frac{d\sigma^{\gamma A \to f X}}{d^2 p_\perp}(\omega, p_\perp, Q^2).
    \end{align}
\end{subequations}

The $n_{\gamma}^{pA}(\omega)$ and $n_{\gamma}^{AA}(\omega)$ are the equivalent incoming photon spectrum generated by the nucleus source that remains intact in the final state. We assume that the photon flux from the proton and nucleus can be described by the classical Weissacker-Williams flux (WW), with a broad spectrum~\cite{Baur:2001jj, Greiner:1992fz, Bertulani:2005ru, Baltz:2007kq, Contreras:2015dqa, Klein:2017nqo, Klein:2020fmr}:
    \begin{eqnarray}
     n^{pA,\, AA}(\omega) = \frac{2 Z_A^2 \alpha_{\text{em}}}{\omega \pi} \left[\zeta K_0(\zeta)K_1(\zeta) - \frac{\zeta^2}{2}\left(K_1^2(\zeta) - K_0^2(\zeta)\right)\right],   
    \end{eqnarray}
where $\zeta = \frac{\omega}{\gamma_L} \left( R_{A} + R_{p, \, A} \right)$, with $\gamma_L = \frac{\sqrt{s_{NN}}}{2m_p}$ being the Lorentz factor. The proton mass, $m_p = 0.938 \, \text{GeV}$, is taken from the PDG~\cite{ParticleDataGroup:2022pth}, and $Z_A$ denotes the charge of the projectile nucleus (atomic number). Then, the integrated cross section for all possible photon energies comes from the integral over $\omega$ of Eq.~\ref{mesonspectrum} and Eq.~\ref{jetspectrum} in the limits $\omega_{\text{max}}$ and $\omega_{\text{min}}$. The minimum photon energy $\omega_{\text{min}}$ for the meson (inclusive single jet) to be produced at the threshold, namely $\left(W^{\text{min}}_{\gamma N}\right)^2 = M_V^2$ (or $\left(W^{\text{min}}_{\gamma N}\right)^2 = M_\perp^2$ for the jet), would be $\omega_{\text{min}}= M_V^2/(4\gamma m_p)$ ($\omega_{\text{min}}= M_\perp^2/(4\gamma m_p)$). For the maximum photon energy $\omega_{\text{max}}$, it carries the entire energy of the parent ion, then $\omega_{\text{max}} = E_N = \sqrt{s_{NN}}/2$.

\subsection{Dipole cross section}

The key component for calculating the transverse momentum spectrum is the dipole cross section, which is determined by the dipole--target scattering amplitude $N$. This quantity has been the focus of extensive research by various groups~\cite{Frankfurt:2022jns}. Over the past decades, several phenomenological models have been developed. These models typically differ in their treatment of the impact parameter dependence and the transition between the linear and non--linear regimes. Currently, the bCGC~\cite{Watt:2007nr} and IP--SAT~\cite{Kowalski:2003hm} models, which rely on distinct assumptions for the treatment of gluon saturation effects, have successfully described the high--precision HERA data for both inclusive and exclusive $ep$ processes. In this work, we use two of these models without impact parameter dependence for the proton case, namely the GBW model~\cite{Golec-Biernat:1998zce, Golec-Biernat:1999qor, Golec-Biernat:2017lfv} and the rcBK model~\cite{Lappi:2013zma} that solves numerically the BK non--linear evolution equations~\cite{Balitsky:1995ub, Kovchegov:1999yj, Kovchegov:1999ua}. 

The GBW model for the dipole cross section is based on a simple saturated ansatz:
\begin{eqnarray}
    \sigma_{\text{dip},\, p}(x, r_\perp) &=& \sigma_0 \left[1 - \exp\left(-\frac{Q_s^2(x)\, r_\perp^2}{4}\right) \right], \quad Q_s^2(x) = Q_{s0}^2 \left(\frac{x_0}{x}\right)^\lambda.
\end{eqnarray}
We adopt the more recent parameter fit from Ref.~\cite{Golec-Biernat:2017lfv}, with the corresponding values listed in Tab.~\ref{tab:gbwparameters}. The saturation scale $Q_s^2(x)$ encodes the onset of nonlinear QCD effects and increases as $x$ decreases, reflecting the growth of gluon densities at small Bjorken-$x$. 

The running-coupling Balitsky--Kovchegov (rcBK) equation governs the small--$x$ evolution of the dipole--proton amplitude and requires an initial condition specified at a starting value $x_0$. We employ the publicly available numerical solution provided by Lappi--Mäntysaari in Ref.~\cite{Lappi:2013zma} distributed in grid form. We have chosen the $\text{MV}^{e}$ setup, whose initial condition fixes the anomalous dimension to $\gamma = 1$ and treats the infrared cutoff $e_c$ as a free parameter, as our baseline configuration. This parametrization provides better control of the infrared cutoff and yields a smoother interpolation between the saturation region at small $k_T$ and the perturbative tail at large $k_T$, which is precisely the kinematic domain most relevant for real--photon photoproduction in UPCs. Moreover, the MV$^{e}$ model gives a better description of the HERA DIS data ($\chi^2/\mathrm{d.o.f.} \approx 1.15$) than the other two setups present in the paper, namely, MV$^{\gamma}$ and MV.

In the Lappi--Mäntysaari solution, the non--perturbative input for the rcBK evolution at the initial value $x_0 = 10^{-2}$ is taken from the McLerran--Venugopalan (MV) model~\cite{McLerran:1993ni}:
\begin{equation}
N(r_\perp)
= 1 - \exp\!\left[
 -\frac{r_\perp^2 Q_{s0}^2}{4}
 \ln\!\left( 
   \frac{1}{r_\perp\,\Lambda_{\mathrm{QCD}}} + e_c \cdot e
 \right)
\right].
\label{eq:MVparam}
\end{equation}
The corresponding initial saturation scale $Q_{s0}^2$ and the infrared cutoff $e_c$ are listed in Tab.~II. The coupling runs according to Balitsky’s prescription~\cite{Balitsky:2006wa, Kovchegov:2006vj}, where $\alpha_s(r_\perp)$ depends on the transverse dipole size. The corresponding parameters can be seen in Tab.~\ref{tab:gbwparameters}.

\begin{table}[t]
\centering
\begin{tabular}{|c||c|c|c|c|c|c|c|c|}
\hline \hline
 & $m_l/$ GeV & $m_c/$ GeV & $m_b/$ GeV & $\sigma_0/$ mb & $\lambda$ & $x_0$ & $e_c$ & $Q_{s0}^2/\text{GeV}^2$  \\
\hline
GBW~\cite{Golec-Biernat:2017lfv} & 0.14 & 1.4 & -- & 27.32 & 0.248 & 0.42 $\times 10^{-4}$ & -- & 1.0  \\
rcBK~\cite{Lappi:2013zma} & 0.14 & -- & -- & 32.72 & -- & 0.01 & 18.9 & 0.060 \\
\hline
\end{tabular}
\caption{Parameters for the GBW and rcBK model used in the numerical calculation. The light quarks refer to $l = d, u$, and $s$.}
\label{tab:gbwparameters}
\end{table}

The fully dipole--nucleus cross section involves the integration over the dipole impact parameter $s_\perp$:
\begin{eqnarray}
    \sigma_{\text{dip},\, A}(x, r_\perp) = 2 \int d^2 s_\perp \, N(x, r_\perp, s_\perp),  
\end{eqnarray}
where $N(x, r_\perp, s_\perp)$ is the dipole--nucleus amplitude and $s_\perp = |\mathbf{s}_\perp|$ is the impact parameter of the center of the dipole relative to the center of the nucleus. The nuclear extension of this amplitude for a large nucleus is made via the Glauber--Gribov approach~\cite{glauber1959lectures, Gribov:1968jf, Gribov:1968gs}, which gives the dipole--nucleus scattering amplitude averaged over all possible configurations of the nucleons in the target nucleus~\cite{Armesto:2002ny}:
  \begin{eqnarray}
    N(x, r_\perp, s_\perp) = 2 \left[1 - \exp\left(-\frac{1}{2}A\ T_A(s_\perp) \sigma_{\text{dip},\, p}(x, r_\perp) \right)\right].
    \label{eq:glaubergribov}
\end{eqnarray}

This approach introduces a dependence on the dipole--nucleus impact parameter $s_\perp$ for the scattering amplitude $N$ by utilizing the usual (integrated) dipole cross section off a proton--target, $\sigma_{\text{dip},\, p}$. The function $T_A(s_\perp)$ (normalized to 1), appearing in Eq.~\ref{eq:glaubergribov}, represents the nuclear thickness function, which is derived from the nuclear density $\rho_A(\mathbf{r} = (\mathbf{s}_\perp, r_3))$, which is well--approximated by the Woods-Saxon distribution~\cite{Woods:1954zz}
    \begin{eqnarray}
    T_A(s_\perp) = \int_{-\infty}^{\infty}dr_3\,\, \rho_{A}(s_\perp, r_3), \quad \rho_{A}(s_\perp, r_3) = \frac{\rho_{0}(A)}{1 + \exp\left(\frac{r(s_\perp, r_3) - R_A}{\delta} \right)}.
    \label{woodsaxon}
    \end{eqnarray}
    Here, $r(s_\perp, r_3) := \sqrt{s_\perp^2 + r_3^2}$ is the distance from the center of the nucleus, $\rho_{0}(A)$ is an appropriate normalization factor dependent on the mass number $A$, $R_A$ is the radius of the nucleus and $\delta = 0.54$ fm is a fitted parameter~\cite{Helenius:2012ny}. 
    
For lumpy nuclei ($A \lesssim 100$), the nucleons are treated as discrete and localized, and the dipole--nucleus interaction must account for event--by--event fluctuations in the positions of the nucleons. Then, we adopt the approach~\cite{Kowalski:2003hm}
  \begin{eqnarray}
    N(x, r_\perp, s_\perp) = 2 \left[1 - \left[1 -\frac{1}{2}\ T_A(s_\perp) \sigma_{\text{dip},\, p}(x, r_\perp) \right]^{A}\right],
    \label{eq:glaubergribovlumpy}
\end{eqnarray}
which smoothly approaches the expression in Eq.~\ref{eq:glaubergribov} in the limit of small $r_\perp$ and large atomic number $A \to \infty$.

\subsection{Fragmentation functions}

To facilitate the detection of experimental signals, it is preferable to reproduce final states that are experimentally accessible. Therefore, we calculated inclusive cross--sections for identified mesons by employing quark fragmentation models. The fragmentation function $D_M^f\!\left(z_f, \mu^2\right)$ denotes the probability for a quark $f$ with longitudinal momentum $z\, q^+$, where $q$ is the four--momentum of the photon, to fragment into a meson $M$ that carries longitudinal momentum $z_f z \, q^+$ and transverse momentum $\mathbf{P}_\perp = z_f\, \mathbf{p}_\perp$, where $\mathbf{p}_\perp$ is the transverse momentum of the quark.

The production of $M$--mesons is then obtained by convoluting the production cross-section of the quark $f$, given by Eq.~\ref{eq:f-jet}, with the non--perturbative fragmentation function $D_M^f$~\cite{Floter:2007xv, Gimeno-Estivill:2025rbw, Goncalves:2025wwt}:
\begin{eqnarray}
       \frac{d\sigma^{\gamma A \rightarrow M X}}{d\omega\, d^2 P_\perp}(x, P_\perp,Q^2)
       &=&
\int_{0}^{1} \frac{dz_f}{z_f^2}\, D_M^f\bigl(z_f, \mu^2) \left[ \int_{0}^1 dz\, \delta(x_f - z x) \left.\frac{d\sigma^{\gamma  A \to fX}}{d^2 p_\perp dz}\right|_{\mathbf{p}_{\perp} = \frac{1}{z_f}\mathbf{P}_{\perp}}\right]. \nonumber \\
& &
\end{eqnarray}
Here, the quark (antiquark) longitudinal momentum fraction is given by $x_f = M_\perp\, e^y/\sqrt{s_{NN}} = 2z\, \omega_{\text{inc}}/ \sqrt{s_{NN}}$. The fragmentation functions $D_M^f$ are normalized such that the sum over all hadron species $M$ satisfies the condition $\sum_M \int_0^1 dz_f\, D_M^f(z_f, \mu^2) = 1$. The differential single jet cross--section is given by Eq.~\ref{eq:f-jet}.

\begin{figure}[!t]
\centering
\subfigure[\label{fig:ratio_pA}]{%
    \includegraphics[width=0.48\textwidth]{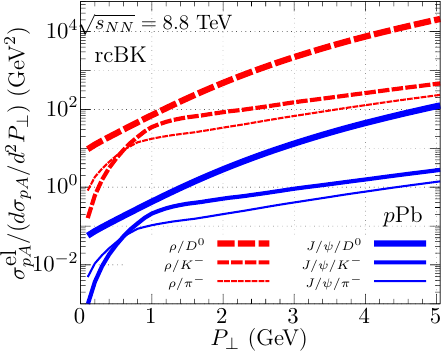}
}
\hfill
\subfigure[\label{fig:ratio_AA}]{%
    \includegraphics[width=0.48\textwidth]{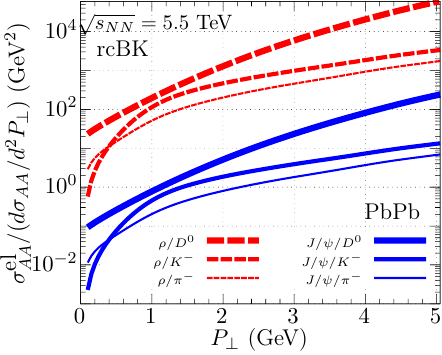}
}
\caption{%
\subref{fig:ratio_pA} The ratio between the cross section for exclusive vector meson photoproduction and the differential cross section on transverse momentum for inclusive meson photoproduction in $p$Pb collisions at the LHC. For the exclusive production, $J/\psi$ (blue lines) and $\rho$ (red lines) mesons are chosen. In the inclusive production, the $\pi^-$, $K^-$, and $D^0$ have been considered. 
\subref{fig:ratio_AA} Same ratio for PbPb collisions at the LHC. The same set of mesons is chosen. In both plots, the results are obtained using the numerical solution of the nonlinear BK equation.%
}
\label{fig:fragmentation_ratios}
\end{figure}

\begin{figure}[!ht]
\includegraphics[width=0.54\linewidth]{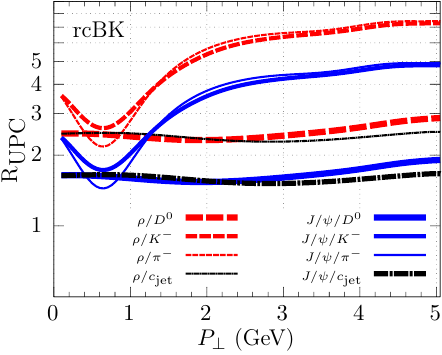}
\caption{The UPC ratio as a function of the open meson transverse momentum (ultra--peripheral $p$Pb and PbPb collisions) for the energies of heavy ion collisions at the LHC (run 4). Same notation as previous figures. The role played by the fragmentation in the case of $D^0$ meson is also presented $(\rho, J/\psi)/D^0$ versus $(\rho, J/\psi)/c_{\text{jet}}$.}
\label{fig:ratio_UPC}
\end{figure}

\section{Results}
\label{Sect:prediction}

In the following, we present our predictions for the $R_{\text{UPC}}$ ratio within the kinematics of LHC Run 4 \cite{Dainese:2019rgk}. We investigate the $A$--dependence of $R_{\text{UPC}}$ and, for the $p_\perp$ distribution, we consider the nuclei probed at the LHC, specifically $p$Pb at $\sqrt{s_{NN}} = 8.8$ TeV, and PbPb at $\sqrt{s_{NN}} = 5.5$ TeV. For our calculations, we assume the light quark masses to be $m_u = m_d = m_s = 0.14$ GeV and the charm quark mass $m_c = 1.4$ GeV, utilizing the GBW and rcBK models. For the $\pi^-$ and $K^-$ mesons, we use the NNPDF parametrization~\cite{Bertone:2018ecm} of the fragmentation functions with QCD evolution, evaluated at the factorization scale given by the transverse jet mass $\mu^2 = M_\perp^2 = P_\perp^2 + m_f^2$. On the other hand, for the $D^0$ meson the LO Kniehl and Kramer parametrization~\cite{Kniehl:2006mw,Bhattacharya:2016jce} is considered:
\begin{equation}
D_M^c(z_c) = \frac{N z_c (1 - z_c)^2}{\left[(1 - z_c)^2 + \epsilon\, z_c \right]^2},
\end{equation}
with $N = 0.577$ and $\epsilon = 0.101$.

We expect that the main observable, $R_{\text{UPC}}$, largely cancels the residual scale uncertainty due to its ratio definition. However, for intermediate observables, our calculation is performed at leading order, so a residual scale dependence may still appear. For this reason, we have adopted $\mu_F^2 = M_\perp^2$, a physically motivated choice that ensures a smooth interpolation between the perturbative and hadron regimes. Within the kinematic range considered ($p_\perp \lesssim 5$~GeV), variations of $\mu_F$ around this central value are not expected to significantly affect the qualitative behavior of the results, as their dominant sensitivity arises from the small--$x$ dynamics encoded in the dipole amplitude rather than from the scale choice.

In the case of jet production, we consider inclusive jet production and include tagging of the final--state particles using fragmentation functions for $D^0$, $\pi^-$, and $K^-$ mesons. Jets are treated purely at parton level (no parton shower, underlying event, or detector-level jet reconstruction is included). For vector meson production, we specifically analyze $J/\psi$ and $\rho$ mesons to study their impact on gluon saturation. Additionally, we investigate the mixed ratio between light mesons and heavy open charm mesons inclusively produced. To calculate the saturation ratio, we assume $R_A \approx (1.2 \,\mathrm{fm}) A^{1/3}$.

\subsection{Transverse momentum dependence}

In Fig.~\ref{fig:ratio_pA}, the ratio between the cross sections for exclusive vector meson photoproduction and the inclusive open meson photoproduction in $p$Pb collisions at the LHC is presented. The ratio is computed as a function of the meson produced in the inclusive process. In the numerical calculation, the solution of the BK equation (rcBK) has been used.  As a representative example of vector mesons exclusively produced, $J/\psi$ (blue lines) and $\rho$ (red lines) have been considered. Moreover, the $\pi^-$, $K^-$, and $D^0$ have been chosen as examples of inclusively produced mesons. As expected, the ratios for $J/\psi$ are smaller than those for the $\rho$ case. For each vector meson, the order of magnitude of the ratio is inversely proportional to the open meson produced. The change of inflection of the ratio in terms of $p_{\perp}$ is directly related to the transverse momentum distribution for the corresponding quark (antiquark), which is subsequently hadronized. Basically, for $p_{\perp}\leq Q_{s}$, the distribution is flattened in comparison with the typical power--like decrease at large transverse momentum $p_{\perp} > Q_{s}$. In $pA$ UPCs, the main contribution comes from the interaction of the photons from the emitting nucleus with the proton target. 

To provide qualitative intuition for the different concavities observed in the $p_\perp$ distributions, we can adopt an illustrative GBW--type estimate. In this picture, the effective momentum fraction at small $p_\perp$ behaves as $x_{\rm eff}\sim (m_f^2+p_\perp^2)/W_{\gamma N}^2$, and is therefore smaller for light than for heavy flavors in the very low-$p_\perp$ region. Since $Q_s^2(x)$ increases as $x$ decreases (e.g.\ $Q_s^2(x)\propto x^{-\lambda}$ in GBW), this implies a larger effective $Q_s(x_{\rm eff})$ for light mesons in that kinematic limit. This qualitative trend is not specific to GBW and is also consistent with our rcBK-based results, although the quantitative magnitude is model--dependent. We further stress that non--perturbative effects and higher--order corrections may affect the size of the effect; however, they are not expected to qualitatively alter the basic kinematic tendency associated with the $x_{\rm eff}$ dependence.


\begin{figure}[!t]
\centering

\subfigure[\label{fig:ratio_pA_atom}]{%
    \includegraphics[width=0.48\textwidth]{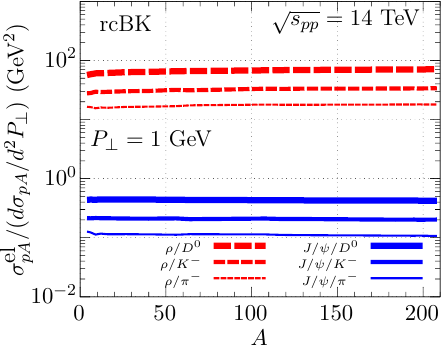}
}
\hfill
\subfigure[\label{fig:ratio_AA_atom}]{%
    \includegraphics[width=0.48\textwidth]{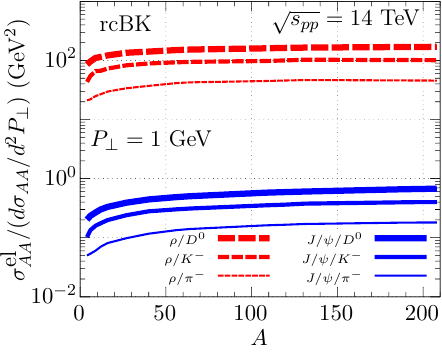}
}
\caption{%
The ratios between the cross sections for exclusive vector meson photoproduction, such as (blue lines for $J/\psi$ and red lines for $\rho$), and those for inclusive meson photoproduction ($\pi^-$, $K^-$, and $D^0$) are shown in \subref{fig:ratio_pA_atom} for $pA$ collisions, and in \subref{fig:ratio_AA_atom} for $AA$ collisions. The transverse momentum is fixed, $P_{\perp} = 1$ GeV.
}
\label{fig:ratio_nuclear_all}
\end{figure}

\begin{figure}[!th]
\centering
\includegraphics[width=0.62\linewidth]{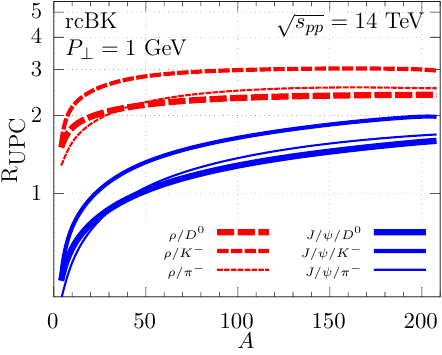}
\caption{The UPC ratio as a function of the atomic number. Same notation as in the previous figure. The transverse momentum is fixed at the value $P_{\perp} = 1$ GeV.}
\label{fig:ratio_rupc1gev_atom}
\end{figure}

In Fig.~\ref{fig:ratio_AA}, the same ratio is shown for PbPb collisions. The pattern is the same as for the $p$Pb case. Notice that in this configuration, the main quantity in the calculation is the $\gamma A$ cross section.
As an effective upper-limit estimate within this simplified baseline, we use the benchmark $Q_{s,\,A}^2(x)\sim A^{1/3}Q_{s,\,p}^2(x)$ at fixed $x$. For lead ($A = 208$), this implies $Q_{s,\,A}/Q_{s,\,p}\sim \sqrt{A^{1/3}}\simeq 2.4$.
However, due to the smaller nucleon--nucleon c.o.m energy of the collision in contrast to the $p$Pb collisions, the values of $x$ probed are higher than in that situation ($\sqrt{s_{NN}}(p\text{Pb})/\sqrt{s_{NN}}(\text{PbPb}) = 1.6$).

The UPC ratio is presented in Fig.~\ref{fig:ratio_UPC}, following the same notation as the previous figures. Moreover, the role of the fragmentation process is investigated by comparing the UPC ratio for the open meson $D^0$ and the corresponding charm jet. The modification is $p_{\perp}$--dependent and alters the ratio; it is more relevant for $p_{\perp}>3$\, GeV. For $D^0$ mesons, $\langle z \rangle (\mu_F = 2m_c)\simeq 0.6$. In the case of heavy open mesons, the UPC ratio is almost flat as a function of $P_{\perp}$. On the other hand, for light mesons, one has a valley structure below $P_{\perp}\simeq 2$ GeV. This can be traced back to a steeper growth of the ratio $\sigma_{pA}^{\text{el}}/(d\sigma_{pA}/d^2P_{\perp})$ on transverse momentum than in the $AA$ collisions.

\subsection{Atomic number dependence}

We now move to the analysis of the atomic number dependence of the distinct ratios. In the numerical calculations, the c.o.m energy of the $pp$ system is set to $\sqrt{s_{pp}} = 14$ TeV. The computation is done for fixed transverse momentum, $P_{\perp} = 1$ GeV. In Fig.~\ref{fig:ratio_AA_atom}, related to $AA$ collisions, the ratio between the cross section for exclusive vector meson production and the differential cross section on $P_{\perp}$ for open mesons is shown. The notation is the same as in previous figures, and the ordinate axis is in logarithmic scale. As expected, the ratio is weakly independent of $A$ (logarithmic growth on $A$), and the highest values are associated with the combination of light vector mesons and a heavy open meson. The same trend is seen in $pA$ collisions; however, in this case, we expect an independent atomic number dependence as the dominant contribution for $pA$ UPCs comes from the photon--proton interaction, where nuclear effects do not play any role.

\begin{figure}[!t]
\centering
\subfigure[\label{fig:ratio_pA_5gev}]{%
    \includegraphics[width=0.48\textwidth]{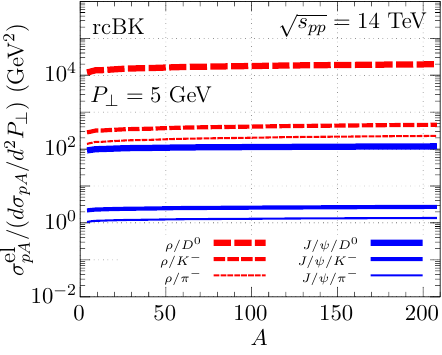}}
\hfill
\subfigure[\label{fig:ratio_AA_5gev}]{%
    \includegraphics[width=0.48\textwidth]{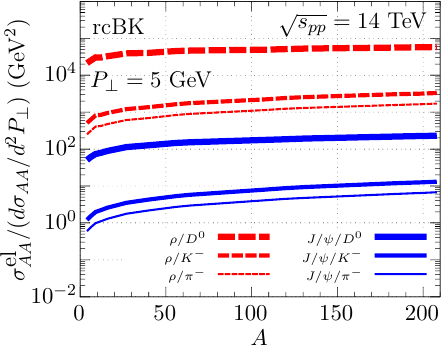}
}
\caption{%
The ratios between the cross sections for exclusive vector meson photoproduction, such as (blue lines for $J/\psi$ and red lines for $\rho$), and those for inclusive meson photoproduction ($ \pi^-$, $K^-$, and $D^0$) are shown in \subref{fig:ratio_pA_5gev} for $pA$ collisions, and in \subref{fig:ratio_AA_5gev} for $AA$ collisions. The transverse momentum is fixed, $P_{\perp} = 5$ GeV.
}
\label{fig:ratio_nuclear_5gev}
\end{figure}

\begin{figure}[!h]
\centering
\includegraphics[width=0.60\linewidth]{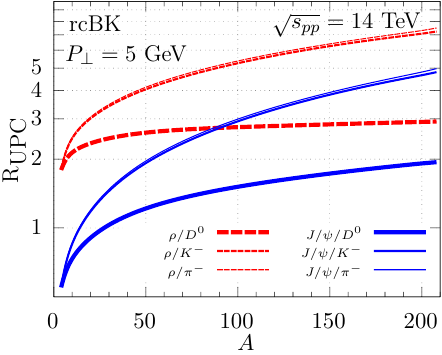}
\caption{The UPC ratio as a function of the atomic number. Same notation as in the previous figure. The transverse momentum is fixed, $P_{\perp} = 5$ GeV.}
\label{fig:ratio_rupc5gev_atom}
\end{figure}

In Fig.~\ref{fig:ratio_rupc1gev_atom}, the corresponding UPC ratio is presented as a function of the atomic number for fixed $P_{\perp} = 1$ GeV. The notation for the curves is the same as in the previous figure. It is clear that the different combinations show a distinct $A$--dependence. We have fitted the UPC ratio with a power--law parametrization, $R_{\rm{UPC}} = C A^{\beta}$, and the effective power $\beta$ has been determined. The corresponding values are shown in Tab.~\ref{tab:fitrupcmesons} for fixed values of transverse momentum $P_{\perp}=1$ and $P_{\perp} = 5$ GeV. In Figs.~\ref{fig:ratio_AA_5gev} and \ref{fig:ratio_pA_5gev} the ratios $\sigma_{AA}^{\text{el}}/d\sigma_{AA}/d^2P_{\perp}$ and $\sigma_{pA}^{\text{el}}/d\sigma_{pA}/d^2P_{\perp}$ are shown for fixed $P_{\perp} = 5$ GeV, respectively. Accordingly, in Fig.~\ref{fig:ratio_rupc5gev_atom}, the corresponding UPC ratio is presented.


\begin{table}[t]
\centering
\begin{tabular}{ |c||c|c|c|c|  }
 \hline
 \hline
              & \multicolumn{2}{|c|}{$P_{\perp}=1$ GeV} & \multicolumn{2}{|c|}{$P_{\perp}=5$ GeV} \\
 \hline
 Configuration & $C$ & $\beta$ & $C$ & $\beta$ \\
 \hline
 $\left.J/\psi \right/ D^0$   & $0.323 \pm 0.006$ & $0.301 \pm 0.004$ & $0.377 \pm 0.007$ & $0.307 \pm 0.004$ \\
 $\left.J/\psi \right/ K^-$   & $0.396 \pm 0.022$ & $0.308 \pm 0.012$ & $0.206 \pm 0.012$ & $0.586 \pm 0.011$ \\
 $\left.J/\psi \right/ \pi^-$ & $0.270 \pm 0.017$ & $0.350 \pm 0.013$ & $0.204 \pm 0.012$ & $0.594 \pm 0.012$ \\
 $\left.\rho \right/ D^0$     & $1.464 \pm 0.052$ & $0.099 \pm 0.008$ & $1.730 \pm 0.059$ & $0.102 \pm 0.008$ \\
 $\left.\rho \right/ K^-$     & $1.718 \pm 0.117$ & $0.116 \pm 0.016$ & $1.087 \pm 0.036$ & $0.351 \pm 0.007$ \\
 $\left.\rho \right/ \pi^-$   & $1.177 \pm 0.078$ & $0.156 \pm 0.015$ & $1.082 \pm 0.037$ & $0.357 \pm 0.007$ \\
 \hline
\end{tabular}
\caption{Result of fit $R_{\rm{UPC}}\propto C A^{\beta}$ for the effective power $\beta$ for the different vector meson-open meson configurations.} 
\label{tab:fitrupcmesons}
\end{table}

Based on Ref.~\cite{Kovchegov:2023bvy}, the following $A$--scaling for $R_{\rm{UPC}}$ is expected for heavy and light vector mesons,
\begin{eqnarray}\label{eq:ratioUPC-theory-H}
    R_{\rm{UPC}}^{\mathrm{heavy}} &\propto &
    \begin{cases}
        A^{\frac{1}{3}\gamma_c}, & p_{\perp}\gg Q_{s},\\
        A^{\frac{2}{3}\gamma_c}, & p_{\perp}\ll Q_{s}.
    \end{cases}\\
    \label{eq:ratioUPC-theory-L}
    R_{\rm{UPC}}^{\mathrm{light}}&\propto &
    \begin{cases}
        A^{-\frac{1}{3}\gamma_c}, & p_{\perp}\gg Q_{s},\\
        A^0 =1, & p_{\perp}\ll Q_{s}. 
    \end{cases}
\end{eqnarray}
where $\gamma_c$ is related to the effective anomalous dimension (critical slope) associated with the small--$x$ evolution. In the simple GBW model, $\gamma_c = 1$, and for the numerical solution of BK, one expects $\gamma_c\sim 0.85$ following the Ref.~\cite{Albacete:2004gw}.  

The double ratio above implies that in the case of the transition from $J/\psi$ to the $\rho$ meson, the powers of $A$ decrease in $R_{\rm{UPC}}$. Therefore, it seems that the ratio is sensitive to how one approaches the saturation regime: (i) lowering $p_{\perp}$ leads to $R_{\rm{UPC}}$ having a higher power of $A$; (i) increasing the average dipole size ($\rho$ meson) leads to a decrease of the ratio for the same values of $A$ and $p_{\perp}$. It should be stressed that $R_{\rm{UPC}}$ behavior on $A$ in Ref.~\cite{Kovchegov:2023bvy} has been obtained assuming a fixed (and similar) value of $x$ in the differential cross section. This is not really the case in $AA$ and $pA$ collisions. In addition, in UPCs actually an average is actually done over the photon energy, meaning that the calculation covers a large range of $x$. This is the main reason for the different $A$--scaling summarized in Tab.~\ref{tab:fitrupcmesons} compared to the estimates in Eqs.~\ref{eq:ratioUPC-theory-H} and \ref{eq:ratioUPC-theory-L}.

\begin{figure}[!ht]
\includegraphics[width=0.54\linewidth]{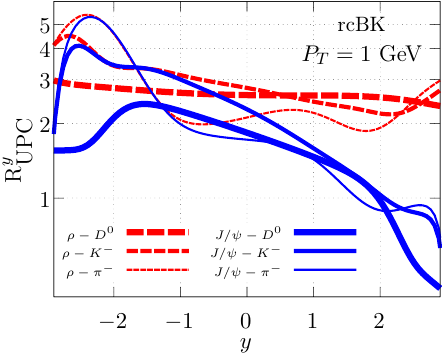}
\caption{The UPC ratio as a function of the rapidity $y$ (ultra--peripheral $p$Pb and PbPb collisions) for the energies of heavy ion collisions at the LHC (run 4). Same notation as previous figures.}
\label{fig:ratio_UPC_y}
\end{figure}

In order to quantify the process of averaging the photon energy, let us consider the relevant photon energy for $AA$ collisions. The rapidity distribution for exclusive processes in UPCs is dominated by the midrapidity contribution, where $W_{\gamma N} (y = 0) = \sqrt{2 M_V m_p \gamma_L}$ and then $x(y = 0) = M_V^2/W_{\gamma N}^2(y = 0)$. This will give $x_{J/\psi}(y = 0)\simeq 6\times 10^{-4}$ and $x_{\rho}(y = 0)\simeq 1.4 \times 10^{-4}$ in PbPb collisions. The saturation scale for protons at this $x$ is $Q_{s,\, p} (x_{J/\psi})\approx 0.72$ GeV and $Q_{s,\, p} (x_{\rho})\approx 0.86$ GeV, respectively. The corresponding nuclear saturation scales are $Q_{s,\, A} (x_{J/\psi})\approx 1.75$ GeV and $Q_{s,\, A} (x_{\rho})\approx 2.1$ GeV. On the other hand, for jets at midrapidities $x_j \approx M_\perp/\sqrt{s_{NN}}$, with $M_\perp^2 = p_{\perp}^2 + m_f^2$ being the jet transverse mass. For jets with $P_{\perp} = 1$ GeV, $Q_{s,\, p} (x_j)\approx 0.8$ GeV and $Q_{s,\, A} (x_j)\approx 2$ GeV. Therefore, even for $P_{\perp} = 1$ GeV, the deep saturation region where $N\rightarrow 1$ is not actually probed. This is the case for very forward rapidities, which marginally contribute to the integrated cross section. Perhaps an alternative definition of the double ratio that circumvents this situation is the following:
\begin{eqnarray}\label{eq:double_ratio_UPC}
    R_{\mathrm{UPC}}^{y} = \frac{\left[d\sigma^{VM}_{\text{el}}/dy/\left(d\sigma^{\text{jet}}_{\text{inc}}/dyd^2 p_{\perp} \right) \right]_{AA}}{\left[d\sigma^{VM}_{\text{el}}/dy/\left(d \sigma^{\text{jet}}_{\text{inc}}/dyd^2 p_{\perp} \right) \right]_{pA}},
\end{eqnarray}

The Fig.~\ref{fig:ratio_UPC_y} shows the $R_{\mathrm{UPC}}^{y}$ ratio defined in Eq.~\eqref{eq:double_ratio_UPC}. Consistently, in the midrapidity region, the curves in Fig.~\ref{fig:ratio_UPC_y} are nearly flat and symmetric around $y=0$, with only a mild rapidity dependence and channel--to--channel variations at the $\mathcal{O}(10\%)$ level (between $J/\psi$ vs.\ $\rho$ and $D^0$, $K^-$, $\pi^-$ references). The bending at larger $|y|$ originates from the asymmetric rapidity dependence in the $pA$ denominator (we only consider the nucleus as photon--emitter), whereas the $AA$ numerator is symmetric for identical nuclei. This is more pronounced for $J/\psi$ than for $\rho$ due to changing $y$ swaps the dominant $\gamma p$ vs.\ $\gamma A$ channel and thus the photon--target energy $W$. Because exclusive $J/\psi$ photoproduction has a much steeper energy dependence, $\sigma_{\gamma p\to J/\psi p}\!\propto\! W^{\delta_{J/\psi}}$ with $\delta_{J/\psi}\!\sim\!0.6$--$0.8$, than light vectors, $\sigma_{\gamma p\to \rho p}\!\propto\! W^{\delta_{\rho}}$ with $\delta_{\rho}\!\sim\!0.2$--$0.3$, the $y$ variation in $pA$ is amplified for $J/\psi$. Equivalently, since $x\!\sim\!M_V^2/W^2$, the heavier $J/\psi$ probes larger effective $x$ and a harder dipole scale (smaller dipoles), making its yield more sensitive to small-$x$ evolution and to traversing distinct nPDF regimes (shadowing $\leftrightarrow$ antishadowing) as $y$ changes.

\begin{figure}[t]
\centering

\subfigure[\label{fig:ratio_pA_jets}]{%
    \includegraphics[width=0.48\textwidth]{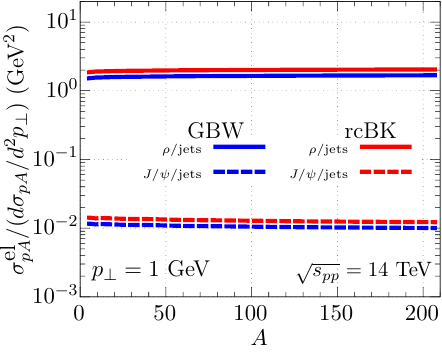}
}
\hfill
\subfigure[\label{fig:ratio_AA_jets}]{%
    \includegraphics[width=0.48\textwidth]{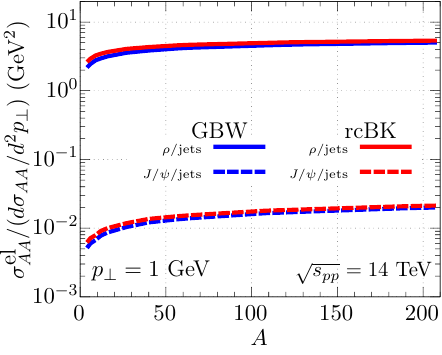}
}

\vspace{0.8em} 

\subfigure[\label{fig:ratio_rupc_jets}]{%
    \includegraphics[width=0.5\textwidth]{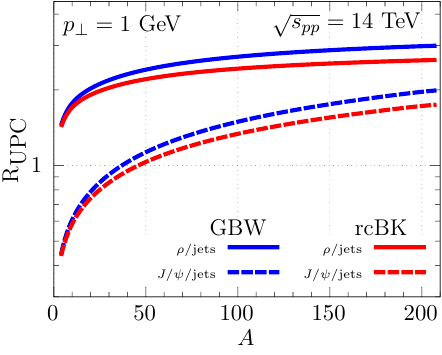}
}

\caption{%
Comparison of the nuclear ratio $\sigma_{\text{el}}/ \frac{d\sigma}{dp_\perp^2}$ for jet production in \subref{fig:ratio_pA_jets} $p$Pb, \subref{fig:ratio_AA_jets} PbPb, and \subref{fig:ratio_rupc_jets} the corresponding UPC ratio. The transverse momentum is fixed, $p_{\perp} = 1$ GeV.
}
\label{fig:ratio_all_jets}
\end{figure}

To understand the effect of the open hadron fragmentation, in Figs.~\ref{fig:ratio_AA_jets} and \ref{fig:ratio_pA_jets}, the ratios are presented for the case of the quark jets instead of open mesons. The theoretical uncertainty coming from the different choices for the color dipole amplitude is also investigated, where the rcBK results are contrasted with those using the GBW model. The small difference is already expected once the effective anomalous dimension is not the same in each case, as mentioned before. For large nuclei, the deviation can reach 25\% and it is consistent with the results for exclusive meson production in UPCs at the LHC~\cite{Goncalves:2017wgg, SampaiodosSantos:2014puz, SampaiodosSantos:2014qtt}.

\begin{figure}[!ht]
\includegraphics[width=0.54\linewidth]{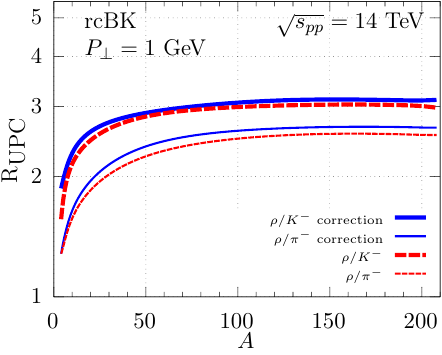}
\caption{Comparison of the UPC ratio, $R_{\text{UPC}}$, as a function of the mass number $A$ for ultra--peripheral $p$Pb and PbPb collisions at LHC (Run 4) energies. The blue lines incorporate the photon wave function nonperturbative correction (Eq.~\ref{eq:wavecorrection}), while the red lines represent the previous calculation.}
\label{fig:ratio_UPC_nonpert}
\end{figure}

To investigate the sensitivity of our results to large--$r_\perp$ non--perturbative modeling, we incorporate the phenomenological ``soft factor'' correction proposed in Refs.~\cite{Fagundes:2024jcq, Goncalves:2020cir} by modifying the photon wave function as
\begin{eqnarray}
 \Psi_{T,L}^f(\mathbf{r}_\perp, z, Q) \to \sqrt{f_s(\mathbf{r}_\perp)}\, \Psi_{T,L}^f(\mathbf{r}_\perp, z, Q), \qquad f_s(\mathbf{r}_\perp) = \frac{1 + B e^{-\omega^2 (r_\perp-R)^2}}{1 + B e^{-\omega^2 R^2}},   
 \label{eq:wavecorrection}
\end{eqnarray}
where $B=-0.75$, $\omega=0.25~\mathrm{GeV}$, and $R=6.8~\mathrm{GeV}^{-1}$. The comparison is shown in Fig.~\ref{fig:ratio_UPC_nonpert}: the blue curves include the correction in Eq.~\eqref{eq:wavecorrection}, while the red curves correspond to our baseline calculation. Within this specific model, the change in $R_{\text{UPC}}$ is small and does not significantly alter the $A$--dependence of the double ratio; the dominant effect is an overall normalization. We stress that this procedure is model--dependent and is intended as a sensitivity test rather than a model--independent quantification of confinement--scale physics.

\section{Conclusions}
\label{Sect:conclusions}

Using the QCD color dipole picture for inclusive jet and exclusive vector meson production in UPCs, we made detailed predictions for the double ratio $R_{\mathrm{UPC}}$ and the ratio $\sigma_{\text{el}}/d\sigma_{\text{jet}/\text{meson}}/d^2p_{\perp}$ in $pA$ and $AA$ collisions as well. We found that $R_{\mathrm{UPC}}\propto A^{\beta}$, with $\beta\sim 0.3--0.5$ for heavy configurations and $\beta\sim 0.1--0.3$ for the light ones. These results are almost independent of the particular choice of the model of the color dipole amplitude. We demonstrated that the $A$-scaling is different from the original proposal in Ref.~\cite{Kovchegov:2023bvy}, especially for light vector mesons. This can be traced back to the integration over photon energies (rapidities of produced particles) in UPCs calculations. We have proposed an alternative version of the double ratio that prevents this shortcoming.

Recently, the full impact-parameter dependence in small-x evolution has been investigated. In Ref. \cite{Mantysaari:2023xcu}, the $b$-dependence is already built into the JIMWLK framework (plus  event-by-event fluctuating proton geometry), and in Refs. \cite{Cepila:2025rkn,Cepila:2025ujv}, the BK equation is solved in the target rapidity and includes the full impact-parameter dependence (dipole orientation). The impact--parameter ($b$) dependence is essential for characterizing spatial gluon density fluctuations, and its omission can artificially enhance saturation effects in proton targets~\cite{Penttala:2024hvp, Mantysaari:2023xcu, Mantysaari:2024zxq}. Moreover,  Ref. \cite{Nemchik:2025myg} shows that a correct incorporation of $r-b$ correlation effects is crucial and may substantially affect the expected signal of gluon saturation.  Despite our coherent observable not explicitly including $b$--dependence, we expect the $b$ sensitivity to partially cancel in the double ratio, ensuring that the overall effect is small and does not alter our qualitative conclusions. The reason is that the ratio is obtained using the $|t|$-integrated cross section for exclusive meson production. On the other hand, the $b$-dependence  of shadowing is important  for the differential cross section $d\sigma_{\gamma A \rightarrow VA}/dt$.

Our analysis is performed at leading order (LO) and should be regarded as a baseline estimate for the proposed observable due to both theoretical and phenomenological limitations. For exclusive vector meson production, NLO calculations are currently restricted to proton targets, preventing a fully consistent implementation in our mixed proton and nuclear study. At the energies considered here, LO and NLO predictions for this process are known to be largely coincident within uncertainties. Regarding single inclusive jet photoproduction, recent NLO developments~\cite{Caucal:2024cdq, Altinoluk:2025dwd} highlight that higher--order effects may become enhanced in very forward rapidity regions, and they are primarily analytical, lacking a comprehensive numerical evaluation. Since our observable is integrated over the experimental rapidity acceptance and constructed as a double ratio, we expect some reduction in sensitivity to such corrections; nevertheless, potentially sizable theoretical uncertainties from missing NLO (and possible resummation) effects remain. A dedicated numerical NLO study will be needed to quantify these effects in a controlled way.

\section*{Acknowledgments}

We would like to thank Emmanuel G. de Oliveira for fruitful discussions. This work was supported by FAPESC, INCT-FNA (464898/2014-5), and the National Council for Scientific and Technological Development – CNPq (Brazil) for JVCL, EH, and MVTM. This study was financed in part by the Coordenação de Aperfeiçoamento de Pessoal de Nível Superior -- Brasil (CAPES) -- Finance Code 001. 

\bibliographystyle{h-physrev}
\bibliography{upcratiobiblio}

\end{document}